\documentstyle[aasms4]{article}
\slugcomment{submitted to AJ, 18 Dec 96}
\lefthead{Zacharias, N.}
\righthead{Astrometric quality of the UCA}
\begin{document}

\title{Astrometric quality of the USNO CCD Astrograph (UCA)}
\author{N. Zacharias\altaffilmark{1} }
\affil{U.S. Naval Observatory, 3450 Mass. Ave. N.W., Washington D.C. 20392,
       nz@pyxis.usno.navy.mil}

\altaffiltext{1}{with Universities Space Research Association (USRA),
  Division of Astronomy and Space Physics, Washington D.C.} 

\begin{abstract}
The USNO 8--inch astrograph has been equipped with a Kodak 1536x1024
pixel CCD since June 1995, operating in a 570--650 nm bandpass.
With 3--minute exposures well exposed images are obtained in the magnitude
range $R \approx 8.5 - 13.5^{m}$.
An astrometric precision of 10 to 15 mas for those stars is estimated from
frame--to--frame comparisons.
External comparisons reveal an accuracy of about 15 mas for those
stars in a 20' field of view.
For fainter stars, the error budget is dominated by the S/N ratio,
reaching $\approx 100$ mas at $R=16^{m}$ under good observing conditions.

\end{abstract}


\section{INTRODUCTION}

Originally the USNO 8--inch Twin Astrograph was equipped with 2 lenses
corrected for the blue and visual bandpasses respectively. 
The instrument has been used for observing the TAC (Twin Astrographic Catalog),
a northern hemisphere photographic catalog 
(\cite{DH90}, \cite{TAC}) and secondary reference stars for the
Radio--Optical Reference Frame (RORF) project (\cite{RORF}) in the
southern hemisphere.
In 1993 the blue lens was replaced by a state--of--the--art 
5--element lens, which is corrected for a 550 to 710 nm red spectral bandpass
(\cite{RLdesign}).
The yellow (visual) lens is now used only for guiding with an SBIG ST4 
autoguider, which moves on an x,y stage in a 2 x 2 degree field of view (FOV).
Since June 1995 a Kodak 1536x1024 pixel CCD camera with a 9 $\mu$m pixel size
has been used as a detector. The option for photographic plates has not been lost
and the CCD camera can easily be exchanged with a plate holder.
This instrument is now called the USNO CCD Astrograph (UCA) and 
currently is located on the Washington DC grounds of the US Naval Observatory.
Table 1 gives details about the telescope optics and Table 2 about the CCD camera.
Currently the telescope is being upgraded to automate the observing 
(\cite{auto}).

Preliminary results of the CCD astrometry achieved with this instrument 
have been reported by (\cite{AAS}) and (\cite{DDA}).
Here, we will investigate the astrometric precision and accuracy in more detail,
including external comparisons.
Section 2 describes the observations and reductions,
results are presented in Section 3,  and Section 4 describes methods of 
quality control for routine operation and future options.

\section{OBSERVATIONS and REDUCTIONS}

\subsection{Focusing}

The focus of the instrument depends mainly on temperature
and temperature gradients, both spatial and temporal.
Particularly during the cooling down period after opening
the dome, frequent determinations of the correct focus 
setting are required for optimal results, because
it cannot be predicted with sufficient accuracy. 
Half a dozen probes measure the air and telescope temperature
at various places. The strongest correlation of focus 
was found with the telescope backend tube temperature.
Focus changes of 0.2 mm/hour have been observed.
After 2 to 3 hours of observing the focus often stays constant even 
if the temperature is still dropping further.
The tolerance in focus setting in good seeing is about 0.05 to 0.10 mm
based on CCD frame reduction results.

Various methods for focusing have been tested.
Finally, a Hartmann screen technique has been adopted for routine use.
A screen with 2 circular apertures of 37 mm in diameter separated by
180 mm is placed in front of the lens.
A single CCD frame is taken with multiple intra-- and extrafocal 10 second
exposures on a 6th magnitude star, shifting the detector columns
between exposures. 
The separation of the "double star" on the resulting frame is highly
correlated with the focus setting. A linear least--squares fit reveals the
location of the intersection of the rays with a formal error of about 0.02 mm.
The location of the holes in the Hartmann screen are chosen to
represent the full aperture "best" focus.

\subsection{Data Acquisition}

Most of the guided CCD frames used for this analysis were taken
with 3 minute exposure times. 
The detector is thermoelectrically cooled to about $-30^{\circ}$C 
and stabilized to within $0.1^{\circ}$C. 
The background is not dominated by dark current but by the bright
Washington DC sky.
Except for some tests, no diffraction grating was used until August 96. 
Most frames were taken within 1.5 hours of the meridian with 
the astrograph on the east side of the pier.
The frames are stored as FITS files on a 386 PC--AT after
a 16 bit A/D conversion, which takes about a minute.

In order to obtain a high star density, the frames for this
analysis were taken close to the galactic plane.
Fields from the radio--optical reference frame (RORF) project (\cite{RORF})
were selected.
For the regular program, a mosaic of 3--by--3 frames shifted
by about 5' was taken centered on each radio source.

\subsection{Raw Image Processing}

Images were transferred to an HP workstation for processing.
A utility program purges the least significant bit of
the pixel data, which is then stored as 2 byte
signed integers with a 15 bit dynamic range.
The IRAF 2.10.4 software was used for the raw data processing.
The object frames were corrected only for dark current,
which includes the bias.
No flat fields were applied to the data.
Preliminary tests with flat fields revealed no significant
advantage for the astrometric results. 
SAOimage was used for image display and visual inspection.

Approximate instrumental magnitudes were determined from the
integral signal as part of the astrometric reduction.
A few photometric standard fields (\cite{Lan}) were taken
to determine the offset between the instrumental and photometric
R magnitudes in order to estimate the limiting magnitude.

\subsection{Astrometric Frame Reduction}

A preliminary version of the Software for Analysing Astrometric CCD's 
(SAAC) (\cite{LarsDiss}) 
was used for the detection of objects and image profile fits.
Image detection is based on a minimum number of 3 consecutive
pixels, each of which exceeds a threshold (S/N=3) above the background.
No attempt was made to identify and retrieve faint stellar images
close to the background level.

Full 2--dimensional Gaussian profiles were used as image models in
nonlinear least--squares adjustments.
A variable aperture size was used for the profile fit, depending on 
the brightness of the star using about 15 to 50 pixels with equal
weights.
Details of the fit algorithm are described elsewhere (\cite{LarsDiss}).
Positions are based on a circular symmetric profile function
(5 parameters).
A second 2--dimensional Gaussian fit was performed
using an elliptical profile with 7 parameters (\cite{Schramm})

\[  I(x,y) \ = \ B \ + \ I_{0} \ 
  \exp \left( \frac{-0.5}{1-\rho^{2}}
   \left[ \left( \frac{x-x_{0}}{r_{x}} \right)^{2} 
        + \left( \frac{y-y_{0}}{r_{y}} \right)^{2} 
	- 2 \rho \left( \frac{x-x_{0}}{r_{x}} \right)
		 \left( \frac{y-y_{0}}{r_{y}} \right) \right] 
       \right) \]

where B = background level, $I_{0}$ = amplitude,
$x_{0}, y_{0}$ = centroid position, $r_{x}, r_{y}$ = radii of the profile
width along the major and minor axes and $\rho$ = orientation. 
The orientation parameter $\rho$ is in the range of +1 to -1
and is the tangent of the angle between the direction of $r_{x}$
and the x--axis. Either $r_{x}$ or $r_{y}$ can be the major axis.
The elliptical fit model is formulated this way because of
a larger numerical stability than is provided by other formulas.
The circular symmetric model is obtained by setting $\rho = 0$
and $r_{x} = r_{y}$.
Results from more advanced image profile modelling will be
presented by (\cite{LarsDiss}).

Double stars were not handled properly at this point.
Large fit residuals or aborted fits were associated with double stars.
All those stars were simply discarded from the following analysis.
Galaxies were handled the same way, although down to
about 15th magnitude there are not many galaxies.

Stars from overlapping frames of each field have been matched
by $x,y$ position only, using the {\em a priori} knowledge of the
approximate frame offsets.
Frame--to--frame transformations were performed on the $x,y$ data
using various models (offset only, orthogonal and full linear, distortion term
etc.) in least--squares adjustments.
Weights were assigned to each stellar image as a function of the
profile fit precision plus a constant global noise added for all images of
a frame, which depends on the exposure time. 
This allows for the effects of atmospheric turbulence on
differential astrometric observations (e.g. \cite{atmlim}).

For an external comparison, some CCD frames at small zenith distances
were reduced to the celestial reference frame (right ascension, declination)
using high precision reference star positions.
A linear model was used for these reductions and no corrections
for refraction were made, because of the smallness of the FOV.
No corrections for differential color dispersion were made because of
the narrow bandpass used and the small zenith distances ($\le 30^{\circ}$)
of the observations.

\section{ASTROMETRIC RESULTS}

\subsection{Image Profile Fit Precision}

The image profile fit x,y--position precision, $\sigma_{pf}$,
is a function of the magnitude of the stars (Fig.~1).
For faint stars the signal--to--noise ratio is the dominating error
contribution. 
There is an asymptotic limit, $\sigma_{pfa}$,
for the fit precision of bright stars.
Saturated (overexposed) star images again display larger $\sigma_{pf}$
simply because the assumed model function no longer resembles 
the observed image profile.
The saturation limit for the 3--minute exposure CCD frame of Fig.~1 is
about $8.5^{m}$ on the instrumental magnitude scale, which is close
to the photometric R system.

The usable dynamic range, from the saturation limit to an
arbitrary limit in $\sigma_{pf}$ = 0.1 pixel is about 7 magnitudes.
The shape and properties of the $\sigma_{pf}$ vs.~magnitude plot
does not change with exposure time, and it even remains the same 
for other telescopes (\cite{RORF}).
A theoretical explanation for this fact is given by 
(\cite{LarsDiss}).

Images with significantly larger $\sigma_{pf}$ than the average
for a given magnitude are from double stars or galaxies or include
defects in their fit area.
Most of these images also show profile widths or image elongation 
significantly larger than the average values for a given frame,
which are well defined for stars in the 8 to 15 magnitude range 
(Fig.~2a and 2b). 
Elongation is defined as the ratio of the 2 axes ($r_{x} / r_{y}$)
describing the image profile width radius in the elliptical Gaussian fit.

The limit in the fit precision for this CCD frame is 
$\sigma_{pfa}$ = 0.014 pixel = 0.13 $\mu$m = 13 mas for a single
star image in each coordinate.
Fig.~3 displays $\sigma_{pfa}$ vs.~mean image elongation for all accepted
CCD frames taken between January and July 1996 for various projects.
The two parameters are clearly correlated and $\sigma_{pfa}$ values
as low as 0.011 pixel = 10 mas are observed.

\subsection{Frame--to--Frame Transformations}

A total of 29 overlapping frames was taken of the radio--optical
reference frame fields 0059+581, 1830+285, and 2201+315 in 3 nights. 
The overlap between two frames ranged from 40 to 90\% and the actual offsets from 
the central field were generated by random numbers for 1830+285 and 2201+315
and a regular mosaic in the case of 0059+581.
Only stars with a profile fit precision of $\sigma_{pf} \le 0.05$ pixels were
selected. The x,y data of each overlap frame were mapped onto the
central frame of each field using a linear transformation model in a weighted 
least--squares adjustment.
The positions, $x_{c}, y_{c}$ and $x_{t}, y_{t}$ from the central 
and the transformed overlapping frames of each star
in the overlap area of those 2 frames were compared. 
The unweighted differences for all stars of all frame pairs in our sample are
displayed in Fig.~4a for the x--axis vs.~instrumental magnitude.
No significant systematic errors as a function of magnitude were yet detected.
An estimate of the standard error, $\sigma_{ff}$,
for a single observation per coordinate 
from the frame--to--frame transformation is derived from these
individual differences, using $\bar{x} = (x_{c} + x_{t})/2$ and n=2 in our case,

\[ \sigma_{ff} \ = \ \sqrt{ \frac{ \sum_{i=1}^{n} {(x_{i} - \bar{x})^{2}}}{n - 1}} 
		 = \ \sqrt{ \frac{ (x_{c} - x_{t})^{2}}{2}}  \]

The data were binned, and averages of $\sigma_{ff}$ are plotted vs.~instrumental 
magnitude in Fig.~4b.
A limit of $\sigma_{ff}$ of 10 mas is found for well exposed stars
in the 9 to 13 magnitude range.
The limit obtained for $\sigma_{ff}$ from frames taken on the same 
field center are sometimes even below 10 mas.
Frames of the same field taken on different nights also
show the same results. As long as the frames are well guided and
in focus, there are no nightly variations in these differential
astrometric observations.

\subsection{Systematic Errors}

No significant systematic errors (exceeding about 10 mas RMS added noise)
have yet been found with good CCD frames.
The radial residuals vs.~magnitude plots show no systematic errors.
Likewise no significant coma term was found in the data so far.
Guiding is a problem at this level of accuracy and causes noticable
elongated images for a significant fraction of the data. 
Only well guided frames were used for this analysis.
A more detailed investigation about possible smaller magnitude--dependent
systematic errors is planned which will require much more data, 
the use of a grating in front of the lens
and a comparison of frames taken with the telescope on different sides of
the pier as well as flipping the orientation of the camera with respect
to the telescope backend.

Fig.~5 shows radial differences plotted vs.~distance from the frame center.
The data were taken from the example explained in the previous section.
There is no indication of a third--order optical distortion term
in any of the data taken so far. This is expected because of the
known small distortion of the lens and the small field of view
used with the CCD.
A detailed field distortion pattern analysis
is not feasible yet because of the lack of a sufficient number
of external reference stars (\cite{FDP}).
Plots of residuals in x and y vs.~x and y of the fields analyzed for this
paper showed no significant field distortion pattern.

\subsection{External Comparison to M44 Praesepe}

A high precision subset was selected from the astrometric standard
field of the star cluster M44 (\cite{M44}, \cite{M44p}).
Two frames of the central section of M44 were taken with the
CCD astrograph (unfortunately over 2 hours from the meridian).
A conventional plate adjustment (CPA) was made with
the CCD frames x,y data using only stars with image profile fit
errors smaller than 0.04 pixels = 36 mas.
The average precision in the x,y coordinates for those stars is 18 mas.
The average mean formal error of the reference stars selected
from the M44 catalog for this adjustment was 25 mas per coordinate
for the epoch of the CCD observations.
Results from the unweighted CPA solutions are summarized in Table 3.
Using the full set of stars, the standard errors of the CPA 
was found to be 41 and 42 mas for the 2 frames respectively.
Assuming the error estimates for the M44 reference stars are correct,
and there are no systematic errors, we derive a mean accuracy 
$\sigma_{ext} = 33$ mas for the UCA observations from this
external comparison.

No corrections for refraction as a function of color were made.
Most of the M44 standard field observations were made in the
blue and visual spectral bandpass and systematic errors as a function
of magnitude and color are suspected in those data 
(Russell, private communication).
Our CCD observations were made in a red spectral bandpass.
The 23 reference stars selected for our example have a color index
in the range B$-$V = 0.21 to 1.08.
The CPA was repeated with a reduced list of 17 stars with
B$-$V = 0.42 to 0.91.
The standard error of the adjustment dropped to just below
39 mas for both CCD frames. This corresponds to an upper limit
of $\sigma_{ext}$ = 30 mas for our CCD observations.

\subsection{External Comparison to FASTT data}

A field of about 20' around 2201+315 was observed in September 1996
with the Flagstaff Astrometric Scanning Transit Telescope (FASTT)
(\cite{fastt}) in 3 nights, providing 3 independent scans.
In the same month 14 acceptable CCD frames were taken with the UCA with 
random overlaps between 40 and 90\% with respect to the central frame.
Only 67 well exposed stars in the magnitude range of V= 9 $-$ 14 were
selected for this comparison ($\sigma_{pf} \le 0.02$ pixel).
The positions of the individual FASTT scans were projected onto a
tangential plane to obtain $\xi, \eta$ coordinates.
An internal precision of $\sigma_{FASTT} = 19$ mas was estimated by a 
linear transformation of the $\xi, \eta$ between individual scans.
An internal precision of $\sigma_{ff} = 14$ mas was estimated from
frame--to--frame transformations between x,y coordinates of individual
UCA frames with the same linear model.
The same linear model was used for unweighted conventional plate adjustments
(CPA) of the UCA frames x,y data with 22 to 43 reference stars per frame 
taken from individual FASTT scans.
Individual results are given in Table 4.
A mean standard error of $\sigma_{CPA} = 25.9$ mas was obtained.
From the scatter of individual $\sigma_{CPA}$ values in Table 4
a formal error of less than 1 mas on the mean $\sigma_{CPA}$ is estimated.
The quadratic sum of the internal errors of both instruments alone accounts
for 24 mas of $\sigma_{CPA}$, leaving an additional noise component
of only about 10 mas for any possible systematic errors in this external
comparison.
Fig.~6 shows the CPA $\Delta \delta$ residuals of all 14 UCA frame
adjustments with reference stars from the third FASTT scan as an example.

\section{DISCUSSION}

\subsection{Why is it so good ?}

Our CCD observations are on the 15 mas level of accuracy.
With the same 2--meter focal length type of instrument about
80 mas was achieved photographically (\cite{phot}).
There are several reasons which explain the improvement in this
somewhat unfair comparison.
Most of all, the photographic plates used in this comparison typically
cover a FOV of several degrees while our CCD is limited to only about 20'.
Thus errors introduced by the turbulence of the atmosphere, the
telescope optics and geometric stability of the detector are expected to be
much larger in the larger FOV, regardless of the type of detector used.
Furthermore, for the above mentioned photographic results grainy
emulsions were used in order to avoid anticipated problems
with the hyping--process. This gives a relatively large image centering error,
and recent results with fine grain emulsions show a dramatic increase in accuracy
for the photographic technique (Winter \& de Vegt, private communication).
Finally there is still no plate measuring machine in operation
which gives an accuracy of 0.15 $\mu$m (15 mas here) over the
entire plate area.

The UCA optics is a 5--element design of the 1990's and probably
the best lens in the world for this type of astrometry.
Observations presented in this paper utilize a tiny spot near the optical
axis in a narrow bandpass. 
Astrometric results on the 0.015 pixel level of precision are
common with modern CCD observations (e.g. \cite{otherCCD}), where
accuracies on the same level have been achieved in narrow field parallax work
as well.
This paper decribes the first step toward wide--field astrograph--type CCD astrometry.

\subsection{Where is the limit ?}

The asymptotic limit in the profile fit error, $\sigma_{pfa}$,
to some extent simply reflects the difference between the
model function and the real data point spread function (PSF).
This can be seen in radial profile plots of the pixel data with the
fit model function overlayed. A more realistic model function
would lead to smaller residuals in the profile fit and thus to
smaller numbers in the centroiding precision error estimate.
Therefore $\sigma_{pfa}$ is only an upper limit for the estimate of the
measuring precision of a star's x,y coordinates, for well exposed stars
the positions are better than its formal error indicates.
This explains why in a frame--to--frame transformation the
repeatability of the observations for centrally overlapping frames,
$\sigma_{ff}$, is sometimes even less than $\sigma_{pfa}$.  
The $\sigma_{ff}$ values already include the error contribution
from the atmospheric turbulence, $\sigma_{atm}$ in addition to the
{\em real} fit precision of the image profiles.
For our exposure time and field size we estimate 
$\sigma_{atm} \approx$ 5 to 10 mas (\cite{atmlim}).
This implies that for centrally overlapping frames the UCA performs 
close to the limit as set by the atmosphere for these
3--minute exposures.

Results from not centrally overlapping frames give a larger
$\sigma_{ff}$. This is an indication of systematic errors
as a function of the location of the images in the focal plane.
This is not surprising on a level of 10 mas, which is only
$0.1 \mu$m in the focal plane.
Inhomogeneities in the filter for example could cause such effects,
as well as tilt of the CCD chip with respect to the focal plane.
More data need to be taken to construct a field distortion
pattern (FDP), but in general a large fraction of such systematic
errors can be removed by empirical calibration.

External comparisons showed the accuracy of the UCA observations
to be about 15 mas for well exposed stars under
optimal conditions (good seeing, close to zenith, 20' FOV).
This is in agreement with the
precision of the x,y data obtained from overlapping frames.
Results obtained in the M44 field do not contradict the FASTT
comparison result when allowing for realistic systematic errors
in the photographic data.
The dynamic range for well exposed stars is about 5 magnitudes
with the UCA. The S/N ratio increases rapidly for fainter stars,
limiting the precision and accuracy for a single observation to
about 100 mas at about $7^{m}$ below the saturation limit.
In the future, comparisons with the Tycho catalog (\cite{tycho})
and the Hamburg observatory data (\cite{LarsDiss}) are planned.

\subsection{Quality Control}

The sampling with about 2 to 3 px/FWHM makes it easy to
identify cosmic ray events, galaxies and most double stars (see Fig.~2a)
with post--fit parameters like image profile width and elongation. 
These parameters also allow for a quantitative quality control of the data. 
Additional software calculates global statistics over each 
CCD frame for mean FWHM and mean elongation of bright but not
saturated star images, limiting magnitude and magnitude at saturation
level.
In a second step plots of these data for several plates,
e.g. the output of a night or year of observing,
can be visualized and tolerances can be set for accepting good data.
An example has been presented in Fig.~3.

Fig.~7a shows the strong correlation between
the limiting magnitude, here defined as the magnitude where
$\sigma_{pf} = 0.1$ pixels, and the mean FWHM. 
On average, good seeing (and focusing) allows going deeper with the 
same exposure time.
A change of the FWHM from 2.0 to 2.5 pixels results in a loss
of 0.5 magnitudes for astrometry of faint stars.
The saturation magnitude of a CCD frame is correlated with
the limiting magnitude (Fig.~7b).
Deeper frames have a saturation limit shifted to fainter magnitudes 
as well but not by the same $\Delta m$.
The dynamic range usable for astrometry is wider for deep frames,
which are those with a smaller FWHM according to Fig.~7a.
No correlation of image elongation or FWHM has been found 
with hour angle.
The orientation of images from frames with a significantly large
mean image elongation was found to be always close to the x--axis
(right ascension).

\subsection{Future Options}

A 4k CCD camera, covering a full square degree, has been obtained
for this telescope.
The planned USNO CCD Astrometric Catalog (UCAC-S) will cover the
entire southern hemisphere in a 2--fold overlap in less than 2 years
(\cite{ucac-s}).
A grating in front of the lens will extend the dynamic range towards
brighter stars to include about 90\% of the Hipparcos stars using
block adjustment techniques (\cite{basim}).
Based on the results of this paper an accuracy of better than 20 mas
is expected for stars in the 6 to 14 magnitude range, with positional
errors increasing to 70 mas for R=16.
Additional long exposure frames in selected fields will allow a
direct tie to the extragalactic reference frame (\cite{iaus179}),
going about $2^{m}$ fainter. Simultaneous observations of the
RORF sources with larger telescopes are highly desirable to
strengthen the tie.

With accuracies on the 15 mas level, parallaxes
of a large number of field stars become an issue.
With 8k CCD's, which are already on the horizon, giving a
$2^{\circ} \times 2^{\circ}$ FOV, a complete hemisphere could be observed
within about 42 nights (5000 frames, 15 frames/hour, 8 hours/night).
Repeating this 5 times a year for 3 years would allow solving
for positions, parallaxes and proper motions of about 10 million stars
at the 10 mas level or below.
This assumes that systematic errors are not larger than with
the currently used small 1k CCD chip. Upcoming tests with the 4k chip
will give much more insight into this issue.

\section{CONCLUSION}

Internal precisions of 10 to 15 mas per coordinate for a single observation 
have been obtained with the UCA instrument.
For the first time a meaningful external accuracy estimate could be
obtained by a comparison with FASTT data. 
Accuracies on the 15 mas level have been found in agreement with the
overall error budget, leaving an additional error contribution of no more 
than about 10 mas for systematic errors.
More data are required to investigate possible systematic errors below that level.
This amazingly good result for a 2--meter focal length telescope 
is partly due to the high quality of the instrument and the narrow spectral
bandpass, but is also a consequence of the small (20') FOV currently used.
Accurate guiding in order to obtain perfectly round images is the biggest
challenge for the observing procedure.
Software for quality control is in place to run a global sky catalog project,
which is planned to start in the southern hemisphere in summer 1997.
A direct link to the Hipparcos stars, as well as extragalactic sources, is possible.
Observations with the UCA can basically replace The Tycho Catalog (\cite{ESAT})
at current epochs. Together with the Tycho, TAC and AC (Astrographic Catalogue)
data proper motions on the 2 mas/yr level could be derived for stars to about
12th magnitude.
All stars of the Guide Star Catalog could be observed with 50 mas accuracy
or better.

\acknowledgments

L.Winter and C.de Vegt, Hamburger Sternwarte, are thanked for providing
a preliminary version of the SAAC program package, as well as R.Stone,
USNO Flagstaff, for providing FASTT data of a test field.
The author wishes to acknowledge  
T.J.~Rafferty, M.E.~Germain and the instrument shop of the USNO
under the supervision of J.~Pohlman for maintaining and upgrading
the instrument, as well as the Astrometry Department under
direction of F.S.~Gauss for support of this project.
T.E.~Corbin, M.E.~Germain and D.~Pascu are thanked for valuable discussions
and comments.
National Optical Astronomy Observatories (NOAO) is acknowledged for IRAF,
Smithonian Astronomical Observatory for SAOimage
and the California Institute of Technology for the pgplot software.

\clearpage

\newpage

\figcaption[zacharias.fig1.ps]
  {Image profile fit precision along the y--axis vs.~instrumental
   magnitude for a single CCD frame taken with the USNO CCD
   astrograph of the field 1830+285 with a 3 minute exposure. 
   The results are from a 2--dimensional Gaussian fit.
   The saturation limit is at about $8.5^{m}$.
   Results along the x--axis are nearly identical.} 

\figcaption[zacharias.fig2.ps]
  {The image profile full width at half maximum, FWHM (a) and
   image elongation (b) is displayed vs.~instrumental magnitude
   for all images of a single CCD frame.
   Single stars form a narrow, well--defined strip, and other objects 
   such as double stars, galaxies or defects are clearly
   separated in these diagrams.}

\figcaption[zacharias.fig3.ps]
  {The asymptotic image profile fit error, $\sigma_{pfa}$, in milli--pixels
   is displayed vs.~the mean image elongation as defined in the text.
   One dot represents one CCD frame.
   Results from CCD frames taken between January and July 1996 are shown.}

\figcaption[zacharias.fig4.ps]
  {Position differences (a) and corresponding standard errors (b)
   from frame--to--frame transformations displayed vs.~instrumental magnitude.
   In this example results from 29 overlapping frames of 3 fields are shown.
   Only stars with a fit error of 0.05 pixels or smaller are used.
   Each plot point is the average of 10 individual differences.}

\figcaption[zacharias.fig5.ps]
  {Radial differences are plotted vs.~distance from the frame center (radius).
   The data were taken from the same frames as the previous figure.
   Each plot point is the average of 25 individual differences.}

\figcaption[zacharias.fig6.ps]
  {Residuals of $\Delta \delta$ (y--coordinate) of conventional plate adjustments 
   of 14 UCA frames reduced with FASTT reference stars are plotted vs.~V magnitude.
   Each plot point is the average of 4 individual residuals.}

\figcaption[zacharias.fig7.ps]
  {The mean FWHM (a) of a CCD frame and the saturation magnitude (b)
   are plotted vs.~the limiting magnitude of each 
   frame for all frames taken with the UCA in 1995.}


\begin{thebibliography}{}

\bibitem[Douglass \& Harrington 1990]{DH90} Douglass,G.G.,
  \& Harrington,R.S. 1990, AJ 100, 1712

\bibitem[ESA 1997]{ESAT} ESA, 1997, The Tycho Catalogue, ESA SP-1200

\bibitem[Gauss et al. 1996]{ucac-s} Gauss,F.S, 
 Zacharias,N., Rafferty,T.J., Germain,M.E., Holdenried,E.R.,
 Pohlman,J.W. \& Zacharias,M.I., 1996, BAAS (in press),
 also http://aries.usno.navy.mil/ad/ad.html

\bibitem[Germain 1996]{auto} Germain,M.E., 1996,
  in proceedings of the Lowell workshop on Small Telescopes,
  to be published on the WWW at www.noao.edu

\bibitem[H{\o}g et al. 1995]{tycho} H{\o}g,E., Bastian,U.,
  Halbwachs,J.L., van Leeuwen,F., Lindegren,L., Makarov,V.V.,
  Pedersen,H., Petersen,C.S., Schwekendiek,P., Wagner,K.
  \& Wicenec,A. 1995, A\&A, 304, 150

\bibitem[Landolt 1992]{Lan} Landolt,A.U. 1992, AJ 104, 340

\bibitem[Monet et al. 1992]{otherCCD} Monet,D.G., Dahn,C.C., Vrba,F.J.,
  Harris,H.C., Pier,J.R., Luginbuhl,C.B. \& Ables,H.D., 1992, AJ 103, 638

\bibitem[Russell 1976]{M44} Russell,J.L. 1976, 
  dissertation, Univ. of Pittsburgh, PA, USA

\bibitem[Russell 1986]{M44p} Russell,J.L. 1986,
  in Astrometric Techniques, IAU Symp.No.109, edited by H.K.Eichhorn \& R.J.Leacock,
  Reidel Publ.Company, Dordrecht, p.697

\bibitem[Schramm 1988]{Schramm} Schramm,K.J. 1988, dissertation Univ.of Hamburg

\bibitem[Stone et al. 1996]{fastt} Stone,R.C., Monet,D.G., Monet,A.K.B.,
 Walker,R.L., Ables,H.D., Bird,A.R. \& Harris,F.H., 1996, AJ, 111, 1721

\bibitem[Vukobratovich et al. 1993]{RLdesign} Vukobratovich,D., Valente,T.,
  Shannon,R.R., Hooker,R. \& Summer,R.E. 1993, Report of the
  Optical Science Center, Univ.of Arizona, Tucson


\bibitem[Winter 1997]{LarsDiss} Winter,L. 1997, dissertation Univ.of Hamburg,
  in preparation

\bibitem[Zacharias 1992]{basim} Zacharias,N. 1992, A\&A 264, 296  

\bibitem[Zacharias et al. 1994]{phot} Zacharias,N., de Vegt,C., Winter,L.,\& Weneit,W. 1994,
   in Astronomy from Wide--Field Imaging, Proceedings of the IAU Sympos. 161, 
   Edited by H.T.MacGillivray, E.B.Thomson, B.M.Lasker, I.N.Reid, D.F.Malin, R.M.West
   \& H.Lorenz, Kluwer Acad. Publ. p.285

\bibitem[Zacharias 1995]{FDP} Zacharias,N. 1995, AJ, 109, 1880

\bibitem[Zacharias 1996a]{DDA} Zacharias,N. 1996a, BAAS 28, 1186 

\bibitem[Zacharias 1996b]{atmlim} Zacharias,N. 1996b, PASP, 108, 1135

\bibitem[Zacharias 1997]{iaus179} Zacharias,N. 1997,
  in New Horizons from Multi--Wavelength Sky Surveys, IAU Symp.No.179,
  edited by B.McLean (in press)

\bibitem[Zacharias \& Rafferty 1995]{AAS} Zacharias,N. \& Rafferty,T.J. 1995,
  BAAS, 27, 1302

\bibitem[Zacharias et al. 1995]{RORF} Zacharias,N., de Vegt,C., Winter,L.
 \& Johnston,K.J. 1995, AJ 110, 3093

\bibitem[Zacharias et al. 1996]{TAC} Zacharias,N., Zacharias,M.I., 
   Douglass,G.G. \& Wycoff,G.L. 1996, AJ 112, 2336 

\end{thebibliography}
\end{document}